\documentclass[sigconf]{acmart}

\usepackage[inline]{enumitem} 
\setlist[itemize]{leftmargin=*, nosep}
\setlist[enumerate]{leftmargin=*,nosep,label=(\arabic*)}
\newcommand{\subhead}[1]{\noindent{\textbf{#1.}}}

 \copyrightyear{2026}
\acmYear{2026}
\setcopyright{cc}
\setcctype{by}
\acmConference[CHI EA '26]{Extended Abstracts of the 2026 CHI Conference on Human Factors in Computing Systems}{April 13--17, 2026}{Barcelona, Spain}
\acmBooktitle{Extended Abstracts of the 2026 CHI Conference on Human Factors in Computing Systems (CHI EA '26), April 13--17, 2026, Barcelona, Spain}
\acmDOI{10.1145/3772363.3798437}
\acmISBN{979-8-4007-2281-3/2026/04}

\title{NeuroWise: A Multi-Agent LLM ``Glass-Box'' System for Practicing Double-Empathy Communication with Autistic Partners}

\author{Albert Tang}
\affiliation{%
  \institution{Marriotts Ridge High School}
  \city{Marriottsville}
  \state{Maryland}
  \country{United States}
}
\email{alberttang2005@gmail.com}

\author{Yifan Mo}
\affiliation{%
  \institution{Vrije Universiteit Amsterdam}
  \city{Amsterdam}
  \country{Netherlands}
}
\email{y.mo@vu.nl}

\author{Jie Li}
\affiliation{%
  \institution{Cake Researcher}
  \city{Delft}
  \country{Netherlands}
}
\email{info@cake-researcher.com}

\author{Yue Su}
\affiliation{%
  \institution{Vrije Universiteit Amsterdam}
  \city{Amsterdam}
  \country{Netherlands}
}
\email{y.su@vu.nl}

\author{Mengyuan Zhang}
\affiliation{%
  \institution{Vrije Universiteit Amsterdam}
  \city{Amsterdam}
  \country{Netherlands}
}
\email{m.zhang@vu.nl}

\author{Sander L. Koole}
\affiliation{%
  \institution{Vrije Universiteit Amsterdam}
  \city{Amsterdam}
  \country{Netherlands}
}
\email{s.l.koole@vu.nl}

\author{Koen Hindriks}
\affiliation{%
  \institution{Vrije Universiteit Amsterdam}
  \city{Amsterdam}
  \country{Netherlands}
}
\email{k.v.hindriks@vu.nl}

\author{Jiahuan Pei}
\authornote{Corresponding author.}
\affiliation{%
  \institution{Vrije Universiteit Amsterdam}
  \city{Amsterdam}
  \country{Netherlands}
}
\email{j.pei2@vu.nl}

\begin{abstract}

The double empathy problem frames communication difficulties between neurodivergent and neurotypical individuals as arising from mutual misunderstanding, yet most interventions focus on autistic individuals. We present NeuroWise, a multi-agent LLM-based coaching system that supports neurotypical users through stress visualization, interpretation of internal experiences, and contextual guidance. In a between-subjects study ($N = 30$), NeuroWise was rated as helpful by all participants and showed a significant condition $\times$ time effect on deficit-based attributions ($p = 0.02$): NeuroWise users reduced deficit framing, while baseline users shifted toward blaming autistic ``deficits'' after difficult interactions. NeuroWise users also completed conversations more efficiently (37\% fewer turns, $p = 0.03$). These findings suggest that AI-based interpretation can support attributional change by helping users recognize communication challenges as mutual.

\end{abstract}



\begin{document}

\maketitle

\begin{figure*}
\centering
\includegraphics[
  width=\textwidth]{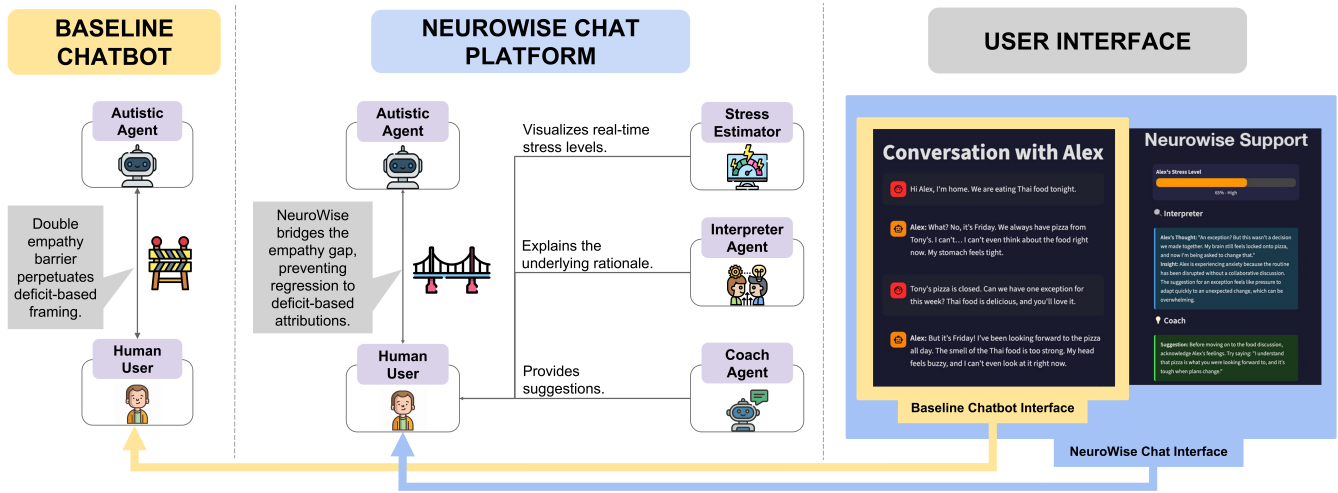}
\caption{Top: Conceptual comparison of Baseline Chatbot versus NeuroWise. Without support, the double empathy barrier perpetuates deficit-based framing; NeuroWise bridges this gap through three agents. Bottom: Example conversation with Alex showing the Baseline (left) and NeuroWise (right) with the support panel displaying stress level, interpretation, and coaching suggestions.}
\Description{Two-part figure. Top: Conceptual diagrams comparing Baseline Chatbot (showing barrier between Autistic Agent and Human User) with NeuroWise Chat Platform (showing bridge connection and three support agents). Bottom: Screenshot mockups of conversations with Alex about routine disruption, with Baseline showing only the chat and NeuroWise showing the chat plus a Support panel with stress level indicator, interpreter explanation, and coach suggestions.}
\label{fig:teaser}
\end{figure*}

\section{Introduction}

Approximately 1 in 36 children in the United States are diagnosed with autism spectrum disorder \cite{maenner2023}, yet communication between neurodivergent and neurotypical individuals remains challenging for both parties. The ``double empathy problem'' \cite{milton2012} suggests that communication breakdowns arise not from deficits in autistic individuals alone, but from bidirectional differences in social cognition and communication styles. While this perspective is now well established, most practical interventions continue to focus on training neurodivergent individuals to adapt to neurotypical norms \cite{crompton2020}. Far less attention has been given to supporting neurotypical individuals in understanding neurodivergent perspectives and adapting their own communication accordingly. Recent work suggests that neurotypical individuals often struggle to interpret autistic communication cues \cite{edey2016} and may unintentionally engage in invalidating or stressful interactions. These difficulties are not simply a matter of lacking communication techniques, but of how breakdowns are interpreted: whether they are seen as personal deficits, situational mismatches, or mutual misunderstandings. However, few tools exist to support neurotypical individuals in developing this kind of understanding through practice.

We present \textbf{NeuroWise}, an AI-powered communication practice environment designed to support neurotypical individuals in interacting more effectively with autistic partners. Rather than focusing solely on improving response strategies, NeuroWise emphasizes helping users understand what may be happening for their autistic partner before deciding how to respond. The system integrates three coaching features: (1) a \textit{Stress Bar} that visualizes the partner's changing emotional state, (2) an \textit{Interpreter} that explains possible internal experiences and triggers, and (3) a \textit{Coach} that provides contextual communication suggestions.

We conducted a between-subjects study ($N=30$) comparing NeuroWise to a baseline chatbot without these features. Our study addresses three research questions: \textbf{RQ1 (Attribution):} How does AI-based interpretation help prevent users from reverting to deficit-based views of autism after difficult communication experiences? \textbf{RQ2 (Acceptance):} How do users perceive AI-supported interpretation and coaching for practicing neurotypical–autistic communication? \textbf{RQ3 (Efficiency):} Do these coaching features support more efficient communication in the simulated interactions?

This paper makes the following contributions: (1) the design of NeuroWise, an AI-powered coaching system featuring stress visualization, interpretation, and communication suggestions for neurotypical-autistic interactions; (2) a multi-agent LLM architecture coordinating autistic partner simulation, stress estimation, interpretation, and coaching in a unified interaction loop; (3) empirical evidence that AI interpretation prevents regression toward deficit-based attributions, directly supporting double empathy theory; (4) design implications for AI-mediated perspective-taking tools that target attribution, not just behavior.

\section{Related Work}

\subsection{Autism Communication and the Double Empathy Problem}

Early research framed autism as a deficit in social communication \cite{apa2013}, with breakdowns commonly attributed to autistic individuals' inherent limitations \cite{anderson2022, kapp2013}. Such attributions reinforce stigma \cite{botha2022} and reduce neurotypical willingness to adapt \cite{sasson2017}. The double empathy problem \cite{milton2012} reframes these difficulties as mutual mismatches in communication styles \cite{crompton2020}. Empirical studies show that non-autistic individuals often struggle to interpret autistic intentions \cite{brewer2016, edey2016, sheppard2016, heasman2019}, while autistic–autistic interactions typically demonstrate smoother communication and mutual understanding \cite{cummins2020, marocchini2024}. Despite these insights, most interventions continue to focus on autistic individuals, with little attention given to supporting neurotypical individuals in understanding and adapting to these mismatches \cite{kapp2019}. NeuroWise addresses this gap by targeting the neurotypical partner.

\subsection{LLM-Based Social Simulation and Coaching}

Large Language Models (LLMs) enable ``generative agents'' that simulate believable, open-ended human behavior \cite{park2023}, extending earlier work on embodied conversational agents \cite{cassell2000}. Such agents have been applied to training contexts including negotiation and conflict resolution \cite{xia2025}. In parallel, LLM-based coaching systems provide users with feedback, suggestions, or revised responses, drawing on established research on the role of feedback in learning \cite{hattie2007, dubiel2025}. 

However, most coaching systems emphasize improving what users say or do next, treating communication breakdowns as problems of execution or skill. In contrast, prior HCI research has shown that how systems help users understand a situation can strongly shape engagement and adaptation. Explanations that clarify intent, context, or underlying conditions support user sensemaking and influence how people interpret breakdowns, shifting reactions away from blame and toward adjustment \cite{liao2020}. Yet this interpretive dimension remains underexplored in LLM-based communication coaching.

Existing AI tools for autism communication, from early virtual environments \cite{parsons2002} to recent LLM-based systems \cite{cao2025, deng2025, ren2023, wang2024}, largely focus on the behavioral training of autistic individuals. NeuroWise addresses this gap by supporting neurotypical users through both simulation and interpretation. Its \textit{Interpreter} component helps users make sense of their partner's internal experiences, while the \textit{Coach} provides just-in-time guidance for responding, enabling users to practice communication grounded in understanding rather than correction.

\section{Design and Procedure}

\textbf{System.} NeuroWise is a web-based practice environment where users engage in text-based conversations with an AI-simulated autistic partner named ``Alex.'' The system is built using Streamlit and OpenAI's GPT-4o-mini model. NeuroWise employs a multi-agent pipeline, inspired by recent multi-agent LLM frameworks \cite{wu2023} (Figure~\ref{fig:teaser}): each user message is processed by a \textit{Stress Estimator} that classifies communication patterns and updates Alex's stress level. When stress increases significantly, the \textit{Interpreter} and \textit{Coach} agents are triggered to provide contextual support.

\noindent
\textbf{Setting: Baseline versus NeuroWise.}
The \textbf{baseline condition} presents the same conversation with Alex but without the Stress Bar, Interpreter, or Coach features, representing a standard chatbot interaction.
The \textbf{NeuroWise} condition includes three coaching features: 
\begin{itemize}
    \item \textbf{Stress Bar:} A visual indicator (0--100\%) showing Alex's current stress level, updated after each message. Drawing on affective computing principles for making emotional states visible \cite{picard1997}. The Stress Bar uses a hybrid algorithm: an LLM (GPT-4o-mini) classifies each user message into communication categories (validation, invalidation, pressure, options-giving, sensory accommodation), and rule-based deltas adjust stress accordingly. The algorithm was pre-validated against two independent human raters on 15 scripted conversations (63 turns). Inter-rater reliability was excellent ($ICC = 0.86, 95\% CI [0.77, 0.91]$) \cite{koo2016}, algorithm-human correlation was strong ($r = 0.86, p < .001$), and discriminant validity between low-stress and high-stress conversations was very large (Cohen's $d = 9.33$). 
    \item \textbf{Interpreter:} When the user's message increases Alex's stress, the Interpreter appears to explain what Alex might be experiencing internally and why, making invisible cognitive and sensory states visible to the user. 
    \item \textbf{Coach:} Provides contextual suggestions for how to respond, focusing on validation, sensory accommodation, and providing options rather than pressure.
\end{itemize}

\noindent
\textbf{Participants.} We recruited 30 neurotypical adults (15 women, 15 men; $M_{\text{age}} = 31.3$, $SD = 7.1$) through online platforms. Contact frequency with autistic individuals varied: 37\% reported high familiarity ($\geq$6 interactions/month), while 63\% reported low-to-moderate familiarity. Participants were randomly assigned to NeuroWise ($n=15$) or Baseline ($n=15$) conditions; groups were balanced on gender and contact frequency (see Supplementary Material).

\noindent
\textbf{Procedure.} Users encounter a scenario designed to elicit common communication challenges: Alex has a Friday pizza night routine that is unexpectedly disrupted when the user arrives home with Thai food instead. This scenario incorporates sensory sensitivities \cite{robertson2017} (strong curry smell), routine dependence and intolerance of uncertainty \cite{wigham2015}, and the need to navigate emotional distress (see Supplementary Material for full scenario instructions).
Participants completed: (1) a pre-survey measuring communication attitudes (8-item double-empathy scale), autism knowledge, and demographics; (2) the conversation task with Alex (typically 8--12 turns); and (3) a post-survey measuring attitude change, perceived learning (5 items), behavioral intentions (3 items), and open-ended reflections. NeuroWise participants additionally rated feature helpfulness (see Supplementary Material for complete survey instruments). Sessions lasted approximately 20 minutes.
Given small sample sizes and ordinal measures, we used non-parametric Mann-Whitney U tests for between-group comparisons and Wilcoxon signed-rank tests for within-group comparisons \cite{romano2006}. Effect sizes are reported using Cliff's delta ($\delta$) \cite{cliff1993}, interpreted as small ($|\delta| < 0.33$), medium ($< 0.474$), or large ($\geq 0.474$).

\noindent
\textbf{Ethical Considerations.}
This study involved human participants and followed applicable institutional ethical guidelines. 
Prior to data collection, the study was assessed using an institutional ethics self-check procedure, 
which determined that a full ethics review was not required. This determination was based on the 
following considerations: (1) participation was voluntary and based on informed consent; (2) no sensitive or identifiable personal data were collected; (3) the study posed no foreseeable risks to participants; (4) participant anonymity and data confidentiality were ensured throughout the study; (5) participants received a modest compensation consistent with standard academic research practices.

\section{Results}

\subsection{Prevention of Deficit-Based Attribution (RQ1)}

Deficit framing was measured with a 2-item composite ($\alpha$ = 0.84) capturing attributional beliefs about responsibility for communication breakdowns. Participants rated their agreement (1--7) with two reverse-scored statements: ``Most communication problems would go away if autistic people changed to match typical social rules'' and ``If my intentions are good, I don't really need to change how I communicate,'' where higher scores indicate stronger deficit-based attribution.

\begin{figure}[h]
\centering
\includegraphics[width=0.9\columnwidth]{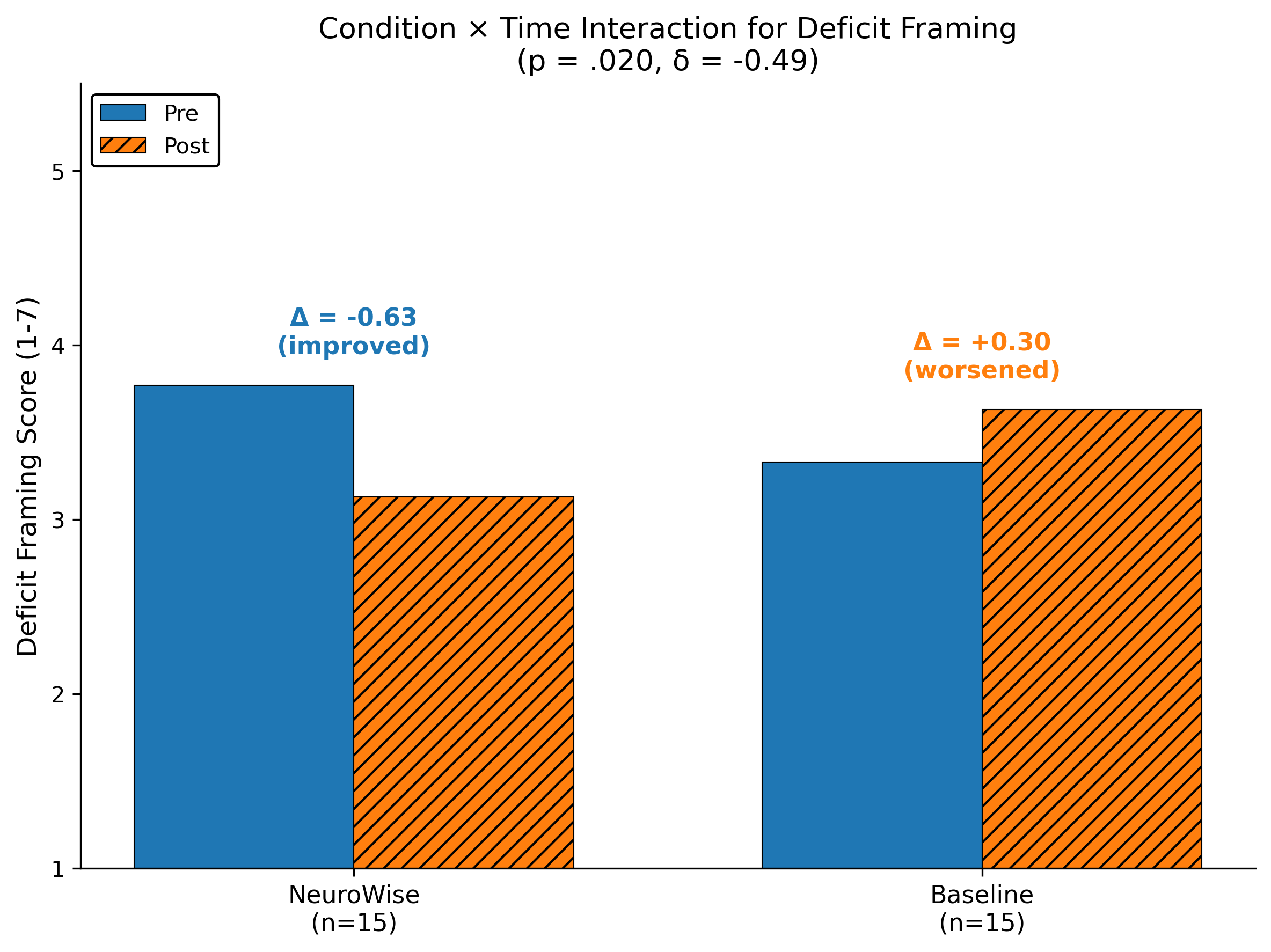}
\caption{Condition $\times$ Time interaction for deficit framing attitudes. NeuroWise participants showed slight improvement while Baseline participants regressed toward deficit framing after difficult conversations.}
\label{fig:deficit}
\Description{Bar chart showing pre- and post-conversation deficit framing scores by condition. NeuroWise shows a decrease from 3.77 to 3.13 (delta = -0.63). Baseline shows an increase from 3.33 to 3.63 (delta = +0.30). The interaction is statistically significant (p = .020, delta = -0.49).}
\end{figure}

We found a significant condition $\times$ time interaction for \textit{deficit framing} attitudes ($U = 57.0$, $p = .020$, $\delta$ = -0.49, large effect; Figure~\ref{fig:deficit}). Neurowise Participants showed a modest reduction in deficit-based framing over time (mean change = -0.63), while Baseline participants shifted in the opposite direction, showing increased endorsement of deficit-based explanations after the interaction (mean change = +0.30). This divergence suggests that, in the absence of interpretive support, challenging communication experiences can lead neurotypical users to increasingly attribute breakdowns to autistic deficits rather than mutual differences. In contrast, interpretive feedback appears to stabilize attribution, helping users resist this regression even when conversations are difficult.

NeuroWise participants also showed significant improvement in \textit{communication flexibility} from pre- to post-survey (Wilcoxon $p = .026$), suggesting the coaching features helped users recognize the need to adapt their own communication style rather than expecting the partner to change.

\subsection{User Acceptance of Coaching Features (RQ2)}

NeuroWise participants reported high perceived usefulness across all system components, with overall helpfulness rated at $Mean$ = 6.60/7 ($SD$ = 0.63). The Interpreter was rated highest ($Mean$ = 6.27, 100\% helpful), followed by the Stress Bar ($Mean$ = 6.00, 93\% helpful) and Coach ($Mean$ = 5.93, 93\% helpful). When asked to compare NeuroWise to generic AI chatbots (e.g., ChatGPT) based on prior experience, NeuroWise participants showed strong preference ($Mean$ = 6.60/7, $SD$ = 0.74), suggesting that participants perceived the system as qualitatively different from general-purpose conversational agents. 

We also conducted a lightweight thematic analysis of open-ended responses. Participants consistently described each feature as supporting a distinct form of sensemaking: the \textit{Stress Bar} made reactions visible (``actually know how my partner reacts {(P04)}''), the \textit{Interpreter} supported understanding by explaining internal states (``made internal states visible, which directly changed how I responded {(P03)}''), and the \textit{Coach} translated that understanding into action (``when I did so, the stress bar went down {(P14)}''). These responses suggest that NeuroWise supported participants in making sense of communication breakdowns and in deciding how to respond.

\subsection{Communication Efficiency (RQ3)}

NeuroWise participants completed conversations in fewer turns than Baseline participants ($Mdn = 8.0 vs 11.0$, $U = 59.0$, $p = .03$, $\delta$ = -0.48, large effect). Despite fewer turns, conversations reached similar resolution states (final stress levels did not differ between conditions, $p = .47$), suggesting NeuroWise improved efficiency without sacrificing outcomes.

\section{Discussion}

\subhead{Theoretical Implications: Supporting Double Empathy}
Our findings support the double empathy perspective, which frames communication breakdowns between neurotypes as mutual mismatches rather than unilateral deficits \cite{milton2012}. Attribution theory suggests that people seek causal explanations when interactions fail, and that these explanations shape subsequent engagement \cite{heider1958, weiner1985}. In our study, participants in the baseline condition increasingly endorsed deficit-based explanations after a difficult interaction, suggesting a tendency to locate responsibility in the autistic partner when no interpretive support was available. In contrast, participants using NeuroWise did not show this shift. The Interpreter component offered contextual explanations for the autistic partner's distress, making the interaction more legible without assigning blame. This interpretive support appears to have helped users maintain a view of the breakdown as a situational and relational challenge rather than evidence of individual deficit. Rather than ``correcting'' users' behavior, the system supported how they made sense of the interaction, highlighting attribution as a key site for intervention in neurotypical–autistic communication.

\subhead{Design Implications}
Our findings suggest that \textbf{supporting understanding of interactions} is more valuable than prescribing correct responses in neurotypical–autistic communication. Participants consistently described the Interpreter as more helpful than direct coaching suggestions. This indicates that perspective-taking and sensemaking are especially important in situations where misunderstandings risk being attributed to personal deficits. \textbf{Visualizing invisible internal states} also emerged as important. The Stress Bar provided participants with a way to monitor how their actions affected their partner over time, without requiring explicit instruction or correction. Rather than telling users when they were ``right'' or ``wrong,'' this form of feedback allowed them to notice patterns and adjust their behavior accordingly. These findings point to the importance of \textbf{designing for attribution, not only behavior}. Systems that support reflection on underlying causes may help users avoid defaulting to deficit-based explanations, even when conversations are difficult. In this sense, AI-mediated communication tools can support understanding and adaptation without reducing interaction to scripted behavior or compliance.

\subhead{Limitations and Future Work}
This study has several limitations. We examined a single interaction scenario focused on routine disruption, and measured immediate effects within a simulated environment rather than real-world interactions. While these constraints allowed controlled comparison, they limit claims about generalization and long-term impact. Future work should explore whether attributional effects persist over time and transfer to everyday communication with real partners. Involving autistic collaborators in the design and evaluation process will be critical, to ensure that interpretive explanations reflect lived experience rather than external assumptions. The partner agent could also be personalized to better reflect diversity across the autism spectrum, including variation in communication profiles, sensory sensitivities, and support needs. Beyond one-directional support, bidirectional tools that help autistic and neurotypical individuals interpret each other's communication patterns may better align with double empathy principles. Future studies could also examine stress-triggered mechanisms, links between observable behavior and subjective ratings, and individual differences such as participants' prior familiarity with autism.

\subhead{AI Usage Disclosure}
We used AI tools in a supportive and limited role. 
Specifically, GPT-5.2 was used for language editing (e.g., improving clarity and conciseness). All findings and figures are based on our own data and results.

\section{Conclusion}

NeuroWise demonstrates that AI-powered coaching can support neurotypical users in developing more effective communication with autistic partners. Our key finding, that NeuroWise prevented regression toward deficit-based thinking while Baseline users regressed, directly supports double empathy theory. By making the partner's internal experiences visible, NeuroWise helped users recognize communication challenges as mutual rather than one-sided. Combined with strong acceptance and efficient communication, AI-mediated interpretation is a promising approach for addressing attributional biases that perpetuate the double empathy gap.

\begin{acks}
We thank the anonymous reviewers for their insightful feedback. We are grateful to colleagues and collaborators for discussions and support that contributed to the design and development of NeuroWise.
We also thanks to the participants who contributed their time and insights to this study.
\end{acks}

\bibliographystyle{ACM-Reference-Format}
\bibliography{references}

\end{document}